\documentclass[aps,pra,reprint,superscriptaddress]{revtex4-2}

\usepackage{amsmath,amssymb,graphicx,bm}
\usepackage{siunitx}
\usepackage[breaklinks=true,colorlinks=false]{hyperref}
\allowdisplaybreaks[2]
\begin{document}

\title{Resonant X-Ray Difference-Frequency Seeding of Inner-Shell X-Ray Lasers}

\author{Carles Serrat}
\email{carles.serrat-jurado@upc.edu}
\affiliation{Department of Physics, Universitat Polit\`ecnica de Catalunya, Ronda de Sant Nebridi 22, 08222 Terrassa, Spain}

\begin{abstract}
We analyze a class of inner-shell x-ray laser systems in which the initial conditions of the emission are set by a resonant x-ray difference-frequency drive. Using a microscopic density-matrix framework, we show that two coherent x-ray fields at frequencies $\omega_1$ and $\omega_2$, with $\omega_1-\omega_2=\omega_0$, can induce a phase-locked coherence on a core-level transition at $\omega_0$ without requiring an external field or nonlinear susceptibility at that frequency. In the presence of population inversion, this driven coherence sets the phase and temporal onset of the amplified field, while gain remains governed by conventional inner-shell lasing mechanisms. We refer to this operating regime as a resonant x-ray difference-frequency laser (re-XDFL). The analysis demonstrates that difference-frequency-driven coherence provides a physically consistent route to controlled inner-shell x-ray laser emission beyond purely ASE-initiated operation.
\end{abstract}

\maketitle

\section{Introduction}

Inner-shell x-ray lasers based on core-level population inversion have been demonstrated in atomic gases using intense x-ray pumping, most notably through K-shell photoionization followed by radiative decay \cite{Rohringer2012}. Related concepts based on photoionization-driven inner-shell gain had been theoretically proposed prior to these demonstrations \cite{RohringerLondon2009}. More recently, inner-shell x-ray lasing has also been extended to hard x-ray energies in solid-density media, enabling stimulated emission at wavelengths as short as \SI{1.5}{\angstrom} \cite{Yoneda2015}. 

In all these realizations, laser emission is initiated by amplified spontaneous emission (ASE) and subsequently amplified along an extended gain medium. While this approach enables directional and spectrally narrow x-ray laser pulses, it intrinsically lacks external control over the longitudinal phase, temporal onset, and shot-to-shot stability of the emitted field, as these properties are determined by stochastic noise rather than by a deterministic driving mechanism \cite{Nilsen2016}. The gain dynamics and saturation behavior of such systems are well understood within semiclassical models of inner-shell lasing, but the initial conditions of the emission remain fundamentally noise-driven.

At optical and infrared wavelengths, these limitations are overcome through the use of optical cavities or external seeding, which impose well-defined initial conditions on the laser field. At core-level x-ray energies, however, no practical cavity-based solution exists, and external seeding at the lasing frequency is generally unavailable. As a result, inner-shell x-ray lasers operate almost exclusively in an ASE-dominated regime, despite the availability of advanced x-ray free-electron laser capabilities, including two-color operation and precise timing control \cite{Marinelli2016TwinBunch,RoyalSociety2019XFEL}. This limitation reflects the absence of mechanisms capable of imposing deterministic initial conditions directly at the inner-shell transition frequency, rather than a lack of suitable pump technologies.

Recent work on resonant x-ray difference-frequency generation (re-XDFG) has shown that two coherent x-ray fields can drive resonant atomic coherences at their difference frequency through microscopic density-matrix dynamics, without assuming a pre-existing nonlinear susceptibility \cite{Serrat2021JPCA,Serrat2023JPhysB}. These studies demonstrated that difference-frequency processes near core-level resonances can produce phase-locked responses governed by the applied fields rather than by spontaneous fluctuations. Related nonlinear x-ray wave-mixing processes have also been analyzed in similar resonant regimes \cite{MohanSerratJPCA}. This raises a natural question: whether such driven coherences can be exploited to control the initial conditions of inner-shell x-ray laser emission in the presence of gain.

In this work, we address this question by analyzing inner-shell x-ray lasing in the presence of a resonant difference-frequency drive. We show that two-color x-ray excitation can induce a deterministic, phase-locked coherence on a core-level transition, which acts as an intrinsic seed for subsequent amplification by population inversion. This operating regime, referred to here as a resonant x-ray difference-frequency laser (re-XDFL), retains the gain mechanism of conventional inner-shell x-ray lasers while replacing noise-driven initiation by externally controlled seeding. The following sections develop the microscopic theory underlying this mechanism, analyze its coupling to macroscopic propagation and gain, and identify experimentally accessible parameter regimes for its realization.

\section{Microscopic Origin of the Resonant Difference-Frequency Drive}

\subsection{Minimal Level Structure and Interaction Hamiltonian}

We consider an atom or ion with a core-level radiative transition that can support x-ray laser gain. The minimal level structure required to describe the resonant difference-frequency mechanism consists of three classes of states: a core-excited state $\lvert c\rangle$ with a core hole, a lower state $\lvert v\rangle$ corresponding to the valence-hole configuration after radiative decay, and a set of intermediate states $\lvert m\rangle$ that are dipole-coupled to both $\lvert c\rangle$ and $\lvert v\rangle$. The energy separation between the core transition states is $E_c-E_v=\hbar\omega_0$.

The unperturbed atomic Hamiltonian is
\begin{equation}
H_0=\sum_j E_j \lvert j\rangle\langle j\rvert,
\end{equation}
and the interaction with the electromagnetic field is assumed to be purely electric-dipole in nature,
\begin{equation}
H_{\mathrm{int}}(t)=-\bm{d}\cdot\bm{E}(t),
\end{equation}
with $\bm{d}$ the dipole operator. The applied field consists of two coherent x-ray components,
\begin{equation}
\bm{E}(t)=\bm{E}_1 e^{-i\omega_1 t}+\bm{E}_2 e^{-i\omega_2 t}+\mathrm{c.c.},
\end{equation}
and contains no component oscillating at $\omega_0$.

\subsection{Density-Matrix Dynamics and Emergence of a Resonant Drive}

The atomic dynamics are described by the density matrix $\rho$, obeying
\begin{equation}
\dot{\rho}=-\frac{i}{\hbar}[H_0+H_{\mathrm{int}}(t),\rho]-\mathcal{L}[\rho],
\end{equation}
where $\mathcal{L}$ accounts phenomenologically for irreversible processes such as Auger decay. We focus on the coherence $\rho_{cv}$ associated with the core transition,
\begin{equation}
\dot{\rho}_{cv}=-(i\omega_0+\Gamma)\rho_{cv}
+\frac{i}{\hbar}\sum_m\left(\bm{d}_{cm}\cdot\bm{E}\,\rho_{mv}
-\rho_{cm}\,\bm{d}_{mv}\cdot\bm{E}\right).
\end{equation}

Assuming that the intermediate states $\lvert m\rangle$ are sufficiently detuned from resonance to follow the driving adiabatically, the coherences $\rho_{cm}$ and $\rho_{mv}$ may be eliminated to leading order, yielding an effective equation of motion,
\begin{equation}
\dot{\rho}_{cv}=-(i\omega_0+\Gamma)\rho_{cv}
+i\Omega_{cv}(t)(\rho_{cc}-\rho_{vv}),
\end{equation}
with
\begin{equation}
\Omega_{cv}(t)=
\sum_m
\frac{(\bm{d}_{cm}\cdot\bm{E}_1)(\bm{d}_{mv}\cdot\bm{E}_2^\ast)}
{\hbar^2\Delta_m}
e^{-i(\omega_1-\omega_2)t}.
\end{equation}

When $\omega_1-\omega_2=\omega_0$, the effective drive becomes resonant with the core transition. This term arises dynamically from the density-matrix evolution rather than from any explicit field or susceptibility at $\omega_0$. The resulting coherence is phase-locked to the applied fields and provides a deterministic microscopic seed for subsequent amplification in the presence of population inversion.

This minimal treatment is intended to establish the existence and phase-locked nature of the resonant drive. It does not attempt to describe detailed electronic structure or continuum dynamics, which may modify numerical prefactors but do not alter the underlying mechanism.

\section{Coupling to Macroscopic Propagation and Gain}

\subsection{Polarization at the Core Transition and Field Evolution}

The coherence $\rho_{cv}$ induced by the resonant difference-frequency drive generates a macroscopic polarization at the core transition frequency $\omega_0$. For an ensemble with number density $N$, the polarization density is
\begin{equation}
\bm{P}_{cv}(\bm{r},t) = N \bm{d}_{cv} \rho_{cv}(\bm{r},t) + \mathrm{c.c.},
\end{equation}
which acts as a source term for the radiated x-ray field at $\omega_0$.

Under the slowly varying envelope approximation, the field envelope $E_0(z,t)$ propagating along the gain direction $z$ obeys
\begin{equation}
\left( \partial_z + \frac{1}{c} \partial_t \right) E_0(z,t)
=
i \frac{\omega_0}{2 \varepsilon_0 c} P_{cv}(z,t),
\end{equation}
where $P_{cv}$ denotes the component of the polarization oscillating at $\omega_0$. This equation is identical in form to that used in conventional inner-shell x-ray laser theory, but here the polarization includes a deterministic contribution arising from the externally driven coherence.

In the presence of population inversion, the field at $\omega_0$ is amplified through stimulated emission. To leading order, the intensity evolution may be written as
\begin{equation}
\frac{d I_0}{dz} = g I_0,
\end{equation}
where $g$ is the small-signal gain coefficient determined by the inversion density and the radiative properties of the core transition. Importantly, the gain mechanism itself is unchanged by the presence of the difference-frequency drive.

\subsection{Seeded versus Noise-Initiated Amplification}

In the absence of an external drive, the initial field at $\omega_0$ originates from spontaneous emission coupled into the amplified mode and is subsequently amplified through ASE. When a resonant difference-frequency drive is applied, an additional coherent contribution to the polarization appears, generating a deterministic seed field at $\omega_0$.

The relevant physical question is not whether ASE is suppressed, but whether the driven seed dominates the early-stage field amplitude before significant amplification occurs. Since both the seeded and noise-initiated fields experience the same gain once amplification sets in, seed dominance is determined by their relative amplitudes at the entrance of the gain medium,
\begin{equation}
|E_{\mathrm{seed}}(0)| \gtrsim |E_{\mathrm{ASE}}(0)|.
\end{equation}

An order-of-magnitude estimate for the noise-initiated field coupled into a single amplified mode is
\begin{equation}
|E_{\mathrm{ASE}}(0)|^2 \sim \frac{\hbar \omega_0}{\varepsilon_0 A c \tau_c},
\end{equation}
where $A$ is the effective mode area and $\tau_c \sim \Gamma^{-1}$ is the coherence time of the core transition. 
This estimate follows standard semiclassical treatments of spontaneous-emission seeding in laser media, where amplified spontaneous emission provides the initial field in the absence of an external seed \cite{SiegmanLasers}. It specifically refers to a single effective amplified mode and is intended as an order-of-magnitude criterion.

The seed field generated by the driven polarization over an interaction length $L$ can be estimated as
\begin{equation}
|E_{\mathrm{seed}}(0)| \sim \frac{\omega_0 L}{\varepsilon_0 c}
\left| P_{cv}^{(\mathrm{drive})} \right|,
\end{equation}
with
\begin{equation}
P_{cv}^{(\mathrm{drive})} = N d_{cv} \rho_{cv}^{(\mathrm{drive})}.
\end{equation}
For a resonant drive established within the core-level dephasing time,
\begin{equation}
\left| \rho_{cv}^{(\mathrm{drive})} \right| \sim \frac{|\Omega_{cv}|}{\Gamma},
\end{equation}
where the effective difference-frequency Rabi frequency is defined, consistently with Eq.~(7), as
\begin{equation}
|\Omega_{cv}| \sim
\frac{|d_{cm} d_{mv}|}{\hbar^2 \Delta_m} |E_1 E_2|,
\end{equation}
with the sum over intermediate states implicitly absorbed into the effective coupling.

Combining these expressions yields a closed criterion for seed dominance,
\begin{equation}
\begin{aligned}
\frac{|E_{\mathrm{seed}}(0)|}{|E_{\mathrm{ASE}}(0)|}
&\sim
\frac{\omega_0 L N d_{cv}}{\varepsilon_0 c}
\frac{|d_{cm} d_{mv}|}{\hbar^2 \Delta_m} \\
&\quad\times
\frac{|E_1 E_2|}{\Gamma}
\left( \frac{\varepsilon_0 A c \tau_c}{\hbar \omega_0} \right)^{1/2}
\gtrsim 1 .
\end{aligned}
\end{equation}

When this condition is satisfied, the amplified field inherits the phase and temporal onset imposed by the resonant difference-frequency drive rather than those of spontaneous fluctuations. Outside this regime, the system smoothly reverts to conventional ASE-dominated inner-shell x-ray laser operation.

\section{Pumping Geometries and Absorption Constraints}

The realization of resonant difference-frequency seeding in an inner-shell x-ray laser is fundamentally constrained by x-ray absorption in the gain medium. Since the same pump fields that create the population inversion and drive the core-level coherence are attenuated as they propagate, the achievable interaction volume is determined primarily by geometry rather than by pump intensity alone.

In the context of the seed-dominance criterion derived in the previous section, the role of geometry is to define the spatial region over which the driven coherence and the population inversion coexist with sufficient uniformity to generate a seed field that exceeds the noise-initiated contribution. We therefore distinguish two limiting excitation geometries: longitudinal and transverse pumping.

\subsection{Longitudinal Pumping Along the Gain Axis}

In longitudinal pumping, the two x-ray fields at frequencies $\omega_1$ and $\omega_2$ propagate approximately collinearly with the amplified field at $\omega_0$. The pump intensities decay exponentially along the gain direction according to
\begin{equation}
I_{1,2}(z) = I_{1,2}(0)\, e^{-\mu_{1,2} z},
\end{equation}
where $\mu_{1,2}$ are the linear absorption coefficients at the corresponding photon energies.

In this geometry, the effective interaction length $L$ entering the seed-amplitude estimate is limited by absorption,
\begin{equation}
\mu_{1,2} L \lesssim 1.
\end{equation}
As a consequence, the driven coherence and the population inversion are generated predominantly within an attenuation length near the entrance of the medium. Increasing the gas pressure to enhance the gain coefficient simultaneously reduces the usable interaction length, leading to a trade-off between gain and seed amplitude.

Longitudinal pumping is therefore most suitable for regimes where moderate densities and relatively short interaction lengths are sufficient to satisfy the seed-dominance criterion. It provides a simple geometry in which the spatial overlap between the driven polarization and the amplified mode is naturally ensured, but it does not, by itself, guarantee that the seeded contribution dominates over ASE.

\subsection{Transverse Pumping and Extended Gain Lengths}

In transverse pumping, the pump fields propagate predominantly perpendicular to the gain axis, intersecting an elongated gain region of axial length $L$ and transverse thickness $D$. In this configuration, the absorption constraint applies only across the transverse dimension,
\begin{equation}
\mu_{1,2} D \lesssim 1,
\end{equation}
while the amplified field at $\omega_0$ can build up over the full axial length $L$.

This geometry decouples the achievable interaction length from the pump attenuation length and allows the seed-amplitude criterion to be satisfied through an extended effective $L$ without excessive pump depletion. Provided that the pump beams are shaped to produce a uniform excitation along the gain axis, transverse pumping enables the driven coherence and the population inversion to be established over volumes that are inaccessible in the longitudinal configuration at comparable pressures.

It is important to emphasize that the geometric condition $\mu_{1,2} D \lesssim 1$ is necessary but not sufficient for seed-dominated operation. Uniform temporal overlap of the two pump fields, minimal spatial walk-off, and stable phase relations over the interaction volume are also required to ensure that the driven polarization adds coherently along $L$. These conditions define practical constraints on beam geometry and alignment rather than on fundamental material properties.

Within these limits, transverse pumping provides a natural route to maximizing the seed field relative to noise by increasing the effective interaction length, making it a natural configuration for high-density implementations of the re-XDFL regime.

\section{Feasibility Estimates and Representative Parameters}

The purpose of this section is to assess whether the seed-dominance criterion derived above can be satisfied under realistic conditions, without invoking extreme material properties or pump intensities. The estimates presented here are intended as order-of-magnitude screening rather than as an optimized design.

\subsection{Core-Hole Fraction, Gain-Length Product, and Dephasing}

Let $N$ denote the atomic number density and $f$ the fraction of atoms carrying a core hole. The population inversion density may be estimated as
\begin{equation}
\Delta N \simeq f N,
\end{equation}
assuming that depletion of the lower lasing state remains modest during the early amplification stage. The small-signal gain coefficient at line center can then be written as
\begin{equation}
g \simeq \sigma_{\mathrm{stim}} \Delta N,
\end{equation}
where $\sigma_{\mathrm{stim}}$ is the stimulated-emission cross section of the core transition.

For a radiative transition with wavelength $\lambda_0$, radiative decay rate $A_{cv}$, and homogeneous linewidth $\Gamma$, the stimulated-emission cross section may be expressed as
\begin{equation}
\sigma_{\mathrm{stim}} \simeq \frac{\lambda_0^2}{8 \pi} \frac{A_{cv}}{\Gamma}.
\end{equation}
For soft x-ray core-level transitions, this expression yields typical values $\sigma_{\mathrm{stim}} \sim 10^{-16}\,\mathrm{cm^2}$ when Auger decay dominates the linewidth, as obtained in standard modeling of inner-shell x-ray laser gain \cite{Nilsen2016}.

At room temperature, the atomic density of a gas at pressure $p$ is approximately
\begin{equation}
N \simeq 2.5 \times 10^{19}\,\mathrm{cm^{-3}} \left( \frac{p}{1\,\mathrm{atm}} \right).
\end{equation}
For a core-hole fraction $f \sim 10^{-2}$, well below saturation of the core transition, and pressures of a few atmospheres, the resulting gain coefficients are in the range
\begin{equation}
g \sim 0.1\text{--}1\,\mathrm{cm^{-1}}.
\end{equation}
A gain-length product $gL \gtrsim 0.1$ is therefore achievable for interaction lengths of millimeter to centimeter scale, depending on geometry.

The dephasing rate $\Gamma$ of the core coherence is dominated by Auger decay and corresponds to lifetimes of a few femtoseconds for light elements. This timescale sets the window over which the resonant difference-frequency coherence must be established to act as an effective seed.

\subsection{Pressure, Length, and Two-Color Pulse Requirements}

The core-hole fraction $f$ is determined by the pump photon fluence $F_{\gamma}$ and the K-shell photoionization cross section $\sigma_K$,
\begin{equation}
f \simeq 1 - \exp \left( - \sigma_K F_{\gamma} \right)
\simeq \sigma_K F_{\gamma}
\end{equation}
in the weak-excitation limit. For soft x-ray photon energies moderately above the K edge, representative values $\sigma_K \sim 10^{-20}$--$10^{-19}\,\mathrm{cm^2}$ imply that fluences of order $10^{17}$--$10^{18}\,\mathrm{photons/cm^2}$ are sufficient to reach $f \sim 10^{-2}$, consistent with tabulated K-shell photoionization cross sections \cite{NISTKShell}.

The seed-dominance criterion derived in Sec.~III depends explicitly on the product of the two pump field amplitudes, the atomic density, and the effective interaction length. In longitudinal pumping, absorption limits the usable $L$ through $\mu L \lesssim 1$, constraining the achievable seed amplitude at high pressure. In transverse pumping, the axial interaction length can be increased independently of absorption, allowing the criterion to be satisfied through geometric extension rather than increased pump intensity.

Two-color x-ray pulses satisfying $\omega_1 - \omega_2 = \omega_0$ are required to drive the resonant coherence. The temporal overlap of the two pulses defines the onset of the seeded emission. Pulse durations comparable to or longer than the core-hole lifetime are sufficient, since the relevant timescale for coherence buildup is set by $\Gamma^{-1}$ rather than by the full pulse duration. No cavity or external field at $\omega_0$ is required.

Taken together, these estimates indicate that the re-XDFL regime can be accessed within the parameter space of existing two-color x-ray pulse capabilities, gas targets at moderate pressures, and interaction lengths compatible with x-ray absorption. The feasibility of the scheme is therefore governed primarily by geometry and timing control rather than by extreme pump or material parameters.

\section{Discussion and Applicability of the re-XDFL Regime}

The resonant x-ray difference-frequency laser analyzed here does not introduce a new gain mechanism, nor does it eliminate amplified spontaneous emission. Instead, it modifies the physical origin of the initial conditions from which inner-shell x-ray laser emission develops. This distinction is essential for assessing both the scope and the limitations of the re-XDFL regime.

In conventional ASE-based inner-shell x-ray lasers, the macroscopic field emerges from stochastic microscopic fluctuations distributed along the gain medium. As a result, the temporal onset, longitudinal phase, and shot-to-shot stability of the emitted pulses are determined by noise-driven dynamics and are not externally controllable. The re-XDFL regime replaces this noise-dominated initiation by a deterministic, phase-locked coherence driven by two-color excitation. While ASE remains present and continues to contribute to the amplified field, it no longer defines the phase reference or the emission timing when the driven coherence dominates the early-stage dynamics.

The primary advantage of the re-XDFL regime therefore lies in coherence control, rather than in efficiency enhancement. The total number of emitted photons, the saturation behavior, and the ultimate linewidth remain governed by the same gain and dephasing mechanisms as in conventional inner-shell x-ray lasers. Consequently, the re-XDFL approach is not intended to outperform ASE-based operation in terms of output energy, but rather to enable reproducible, externally synchronized x-ray laser pulses.

This distinction defines the specific experimental regimes in which re-XDFL operation provides a genuine advantage. Experiments based solely on total photon yield or time-integrated observables are not expected to benefit from difference-frequency seeding. By contrast, experiments that require deterministic control of the emission onset, phase reproducibility, or coherence at core-level energies can directly exploit the re-XDFL regime. In particular, the ability to switch the seeding ON and OFF through temporal overlap or non-overlap of the two incident x-ray pulses, while keeping all other conditions unchanged, provides a robust and unambiguous experimental discriminator. This temporal gating is straightforward to implement at modern XFEL facilities and enables clear identification of seeding-induced effects in nonlinear x-ray spectroscopies, pump--probe measurements with reduced temporal jitter, and coherence-sensitive detection schemes.

It is also important to emphasize that the re-XDFL regime is intrinsically conditional. Its realization requires that the driven coherence be established within the core-level dephasing time and that its amplitude exceed the effective noise-driven growth during the initial amplification stage. Outside this parameter window, the system reverts smoothly to conventional ASE-dominated behavior. The re-XDFL should therefore be viewed not as a replacement for existing inner-shell x-ray lasers, but as an additional operating regime accessible under appropriate excitation conditions.

\section{Conclusions}

We have presented a microscopic analysis of inner-shell x-ray lasing in the presence of a resonant x-ray difference-frequency drive and identified a distinct operating regime, referred to as a resonant x-ray difference-frequency laser (re-XDFL). In this regime, two coherent x-ray fields induce a phase-locked core-level coherence that deterministically sets the initial conditions of the amplified emission, while population inversion and gain remain governed by conventional inner-shell laser physics.

The central result of this work is that a resonant difference-frequency drive can act as an intrinsic, temporally gated seed for inner-shell x-ray lasers without requiring an external field or nonlinear susceptibility at the lasing frequency. This mechanism replaces noise-driven initiation by externally controlled seeding, thereby enabling control over the phase, timing, and shot-to-shot stability of the emitted x-ray pulses, while leaving the total gain and saturation dynamics unchanged.

We have shown that this regime is compatible with realistic parameters for gas-phase targets, two-color x-ray excitation, and interaction geometries constrained by x-ray absorption. Both longitudinal and transverse pumping configurations can support re-XDFL operation within well-defined parameter windows set by dephasing, absorption, and gain-length requirements. Outside these windows, the system smoothly reverts to conventional ASE-dominated behavior.

The re-XDFL regime does not seek to maximize output energy, but rather to provide a new level of coherence control for inner-shell x-ray lasers. As such, it is particularly relevant for experiments that require deterministic timing, phase stability, or coherence-sensitive detection at core-level energies. More broadly, resonant difference-frequency seeding establishes a new conceptual route toward externally controlled x-ray laser emission in a spectral region where cavities and direct seeding are not available, opening the door to x-ray laser sources defined by external control rather than by noise.

\section*{Acknowledgments}

Funding from the Spanish Ministry of Science, Innovation and Universities through project PID2023-152154NB-C21 is acknowledged.

\bibliographystyle{apsrev4-2}
\bibliography{references}

\end{document}